\documentclass[twocolumn,trackchange]{aastex62}

\graphicspath{{./}{figures/}}
\accepted{for publication in ApJ}
\shorttitle{UV continuum slopes of galaxies using UVIT, HST, KPNO}
\shortauthors{Mondal et al.}

\begin{document}

\title{Observed UV continuum slopes ($\beta$) of galaxies at \textit{z} = 0.40 $-$ 0.75 in the GOODS-north field}

\author[0000-0003-4531-0945]{Chayan Mondal}
\affiliation{Inter-University Centre for Astronomy and Astrophysics, Ganeshkhind, Post Bag 4, Pune 411007, India}

\email{chayanm@iucaa.in, mondalchayan1991@gmail.com}

\author[0000-0002-8768-9298]{Kanak Saha}
\affiliation{Inter-University Centre for Astronomy and Astrophysics, Ganeshkhind, Post Bag 4, Pune 411007, India}

\author[0000-0001-8156-6281]{Rogier A. Windhorst}
\affiliation{School of Earth \& Space Exploration, Arizona State University, Tempe, AZ 85287-1404, USA}

\author[0000-0003-1268-5230]{Rolf A.~Jansen}
\affiliation{School of Earth \& Space Exploration, Arizona State University, Tempe, AZ 85287-1404, USA}

\begin{abstract}
We estimate the UV continuum slope ($\beta$) of 465 galaxies (with luminosities of 0.028 $-$ 3.3 $L^{*}_{z=0.5}$) in the Great Observatories Origins Survey (GOODS) Northern field in the redshift range $z=0.40 - 0.75$\,. We use two AstroSat/UVIT (N242W, N245M), two HST (F275W, F336W), and a KPNO (U) bands to sample the UV continuum slope of selected galaxies between 1215 and 2600~\AA~. The mean (median) and 1$\sigma$ scatter in the observed $\beta$ are found to be $-1.33\pm0.07~(-1.32)$ and 0.60 within the considered redshift range. We do not find any significant evolution in the mean $\beta$ within our redshift window. Our measurements add new data points to the global $\beta$ - $z$ relation in the least-explored redshift regime, further reinforcing the gradual reddening of galaxy UV continuum with cosmic time. We notice no strong consistent trend between $\beta$ and M$_{1500}$ for the entire luminosity range $-21$ $< M_{1500} <-15$~mag. Although, the majority of the most luminous galaxies (M$_{1500} <-19$ mag) are found to have relatively redder slopes. Using UVIT, we detect galaxies as faint as M$_{1500} = -15.6$ mag (i.e., 0.028 $L^{*}_{z=0.5}$). The faintest galaxies (M$_{1500} > -16$ mag) tend to be redder, which indicates they were less actively forming stars during this cosmic time interval. Our study highlights the unique capability of UVIT near-UV imaging to characterize the rest-frame far-UV properties of galaxies at redshift $z \sim 0.5$\,.

\end{abstract}


\section{Introduction}
\label{s_intro}
Emission in the ultra-violet (UV) bands carries important clues about the evolutionary phase of a galaxy. The UV continuum flux, which is sensitive to both the properties of dust and stellar populations in galaxies, follows a power law of the form F$_{\lambda} \propto \lambda^{\beta}$, where slope $\beta$ characterizes the shape of the continuum. As massive young stars (age $<$ 100 Myr) contribute the bulk of the UV photons \citep{kennicutt2012}, one can characterize the star-forming mode of a galaxy from the UV flux. UV photons do suffer high-extinction due to the presence of inter-stellar dust in galaxies \citep{calzetti1994}. Therefore, the rest-frame UV continuum slope (i.e., $\beta$) is a powerful tool to probe both dust and stellar population age of galaxies at different redshifts. The value of $\beta$ is also sensitive to galaxy metallicity \citep{calabro2021}. But among different factors, dust-extinction is likely the dominant parameter that decides the value of $\beta$ \citep{wilkins2013,castellano2014}, followed by stellar population age.

Several studies have empirically modelled extinction laws in local galaxies including the Milky Way \citep{fitzpatrick1999,calzetti1994,gordon2003}. All these laws show a steep rise of the extinction coefficient toward shorter UV wavelengths ($\lambda <$ 1700~\AA~). In the case of distant galaxies, where resolving individual stars is not feasible, the UV continuum shape plays the most crucial role to determine their extinction properties. A relation between the ratio of Far-infrared (FIR) to UV luminosity (L$_{IR}$/L$_{UV}$) and the UV continuum slope has been framed to indicate how $\beta$ depends on extinction at lower \citep{meurer1995,meurer1999} and higher \citep{reddy2012} redshifts. Although, at very high redshift (z $\sim$ 7), \citet{wilkins2013} have cautioned on the applicability of such conversion relation. The measurement of extinction at higher redshift is necessary to estimate the cosmic star formation rate density (SFRD) from the dust-corrected UV luminosity. As the high-redshift galaxies are generally fainter, conducting a large-scale spectroscopic survey to understand $\beta$ is prohibitively expensive. Hence, a deep multi-band photometric survey is the best way to study rest-frame UV continua and understand the extinction properties of faint distant galaxies. 

Deep field observations, so far taken with the Hubble Space Telescope (HST) and Spitzer, have revealed the UV continuum slope of a large number of high-redshift galaxies \citep{wilkins2016,reddy2018,kurcz2014,dunlop2013,jiang2020,hathi2013,finkelstein2012,castellano2012,bhatawdekar2021,bouwens2009,bouwens2012,bouwens2014}. Many of these studies reported specific relations between $\beta$ and redshift and $\beta$ and absolute UV magnitude (M$_{UV}$). Galaxies are found to become redder with decreasing redshift which signifies a gradual increase in dust extinction \citep{bouwens2009,bouwens2014}. The massive galaxies, that are more efficient in star formation and hence dust production, are also found to be redder compared to the less-massive ones \citep{meurer1999,bouwens2009,bouwens2012}. From different studies, it is almost evident that with time, galaxies produce more dust which enhances extinction. \citet{bhatawdekar2021,castellano2012} noticed a relation between star formation rate (SFR) and $\beta$ as well. They reported galaxies with higher SFR to be redder.

\begin{figure}
    \centering
    \includegraphics[width=3.5in]{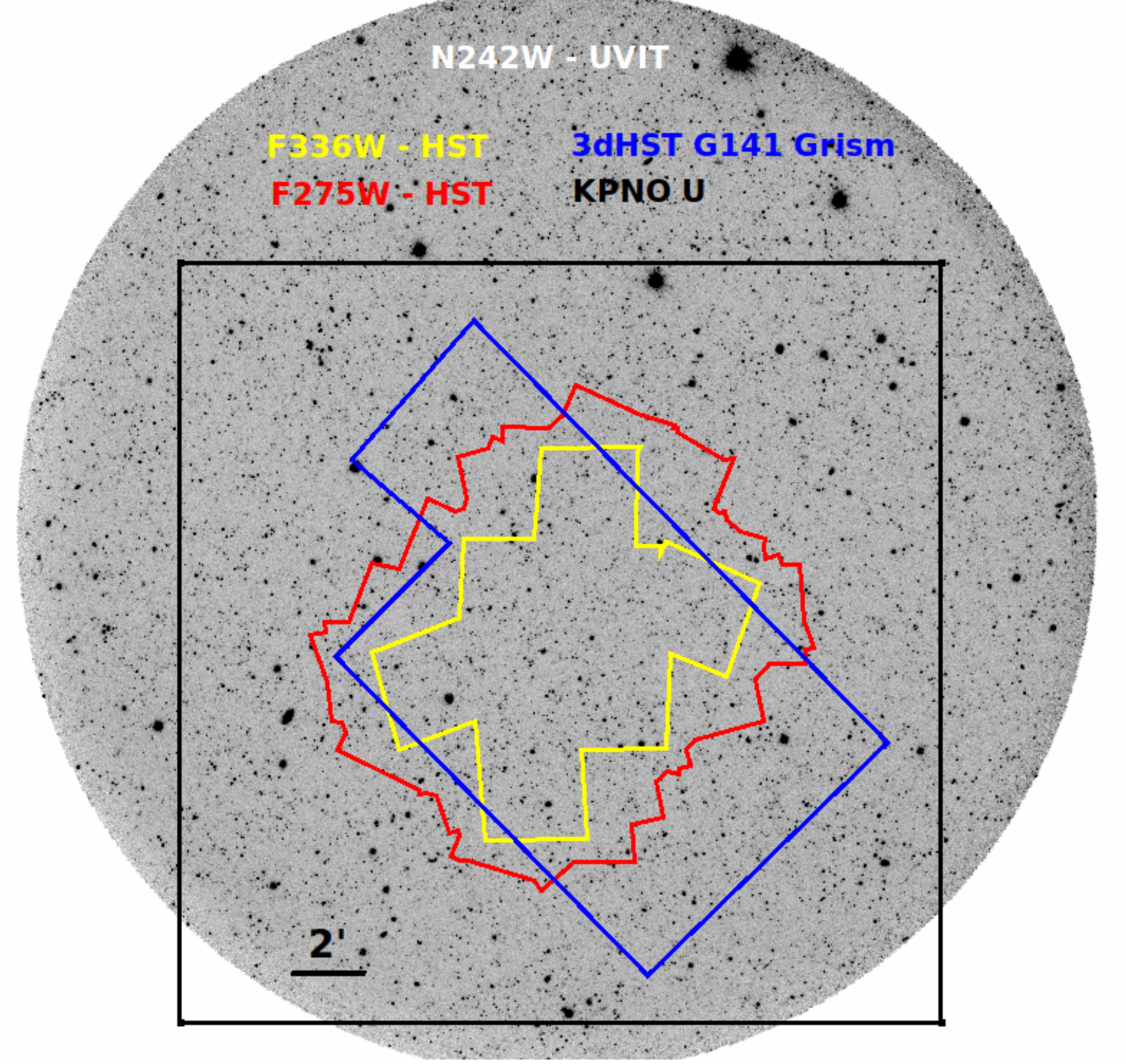} 
    \caption{UVIT N242W band image of the GOODS-north (i.e., AUDFn; \citet{mondal2023}) field. The blue polygon shows the region covered by HST’s WFC3/IR G141 grism observations (AGHAST survey; PI - B. Weiner). The red and yellow polygons represent regions covered by HST F275W and F336W filters respectively (HDUV survey; \citet{oesch2010}). The region covered with KPNO U band is shown in black rectangle. The galaxy sample analysed in this work resides within the blue polygon, where it has been covered by the HST spectroscopic observation.}
    \label{beta_map}
\end{figure}

HST observations have probed $\beta$ of galaxies mostly beyond redshift $\sim$ 1.0, whereas Spitzer IRAC1, IRAC2 bands can sample part of the rest-frame UV continuum for redshift $\gtrsim$ 9.5 \citep{kurcz2014,wilkins2016}. \citet{bouwens2009,bouwens2012,bouwens2014} used HST multi-band observations to study a large number of galaxies at redshifts $\sim$ 2 $-$ 8 in several HST deep fields (for example, HDF-N, HDF-S, GOODS-north, GOODS-south, HUDF/XDF) including five lensing clusters. They reported a clear trend between $\beta$ and M$_{UV}$, where galaxies gradually become bluer with decreasing luminosity. Their study claimed dust as the primary factor that decides the shape of the UV continuum. Although, \citet{finkelstein2012,dunlop2013,bhatawdekar2021} found no significant trend between $\beta$ and M$_{UV}$ for galaxies of different luminosity covering a redshift range $\sim$ 4 - 9\,. However, \citet{finkelstein2012} found the mean $\beta$ to evolve from $-1.82$ to $-2.37$ between redshift of $\sim$ 4 and $\sim$ 7 for their study on the HST deep fields. \citet{wilkins2016} combined both HST and Spitzer data to estimate $\beta$ of six selected galaxies between redshift of 9.5 and 10.5 and found the mean to be $\sim -$2.1\,. This study provides one of the highest redshift samples for $\beta$ which indicated the presence of dust in galaxies even at $z \sim$ 10\,. The majority of the studies reported observed $\beta$ to be redder than the bluest value allowed by the galaxy models (i.e., $\beta$ = $-3.0$). Recently, \citet{jiang2020} used HST and Spitzer data to study six luminous Lyman Alpha Emitters (LAEs) between redshift of 5.7 and 6.6\,. They reported extremely blue values of $\beta$ ranging between $-2.6$ and $-3.4$, although they could not reproduce a galaxy with $\beta = -3.4$ using models. Therefore, the studies reported in the intermediate and high redshift galaxies suggest that the slope $\beta$ can depend on different factors among which dust plays a crucial and dominant role.

In the lower redshift range (z $\lesssim$ 3), there are fewer studies that measured $\beta$. \citet{kurcz2014} used HST/UVIS and HST/WFC3/IR observations to select 923 galaxies spanning a wide redshift range $z \sim 1 - 8$ in the Hubble Ultra-deep Field (HUDF) and found a modest evolution of $\beta$ with both M$_{UV}$ and redshift. They reported the mean $\beta$ value as $-1.382$ for the redshift range {\it z} $= 1 - 2$\,. \citet{reddy2018,hathi2013} also used HST/UVIS filters to probe $\beta$ of galaxies at 1 $< z <$ 3. \citet{hathi2013} noticed the Lyman Break Galaxies (LBGs) at this redshift as more massive and dust-rich (median $\beta = -1.59$ at {\it z} $\sim$ 1.6) than those reported in $z >$ 3. \citet{reddy2018} analysed 3545 star-forming galaxies in the GOODS fields between redshift of 1.5 and 2.5 having a luminosity range $-22.5 < M_{1600} < -17.4$ mag. They found the observed $\beta$ to range from $-2.55$ to 1.05 with a mean $\sim -1.71$\,. \citet{overzier2011} selected a sample of nearby starburst galaxies between redshift of $\sim$ 0.1 and 0.3 and found $\langle\beta\rangle \sim-1.50$\,. \citet{kurcz2014} used GALEX UV photometry for a sample of local dwarf galaxies from \citet{hunter2010} and calculated a mean $\beta$ value of $-1.15$, which is redder than any other measurements at $z > 0$\,. Therefore, the mean $\beta$ measured in these studies suggests that galaxies at lower redshifts are redder which plausibly signifies their higher dust extinction compared to the high redshift galaxies. However, as the $\beta$ value also depends on the physical condition of galaxies, one must be careful while comparing mean $\beta$ from two different studies that sample galaxies of two different properties.

Though, several studies have targeted galaxies at $z \gtrsim$1 (above references) as well as the local dwarfs \citep{hunter2010}, the UV continuum slope has not been explored well at redshift $\sim$ 0.5\,. The primary reason being the lack of high-resolution near-UV (NUV) observation which can sample the rest-frame UV continuum of $z \sim$ 0.5 galaxies. Here, we use the NUV observations taken with the Ultra-Violet Imaging Telescope (UVIT) \citep{kumar2012,tandon2017} to remedy this situation. The NUV bandpass of UVIT has an effective wavelength of $\sim$ 2418~\AA~, which can probe the rest-frame far-UV (FUV) continuum of galaxies at redshift $\sim$ 0.5\,. Apart from the ideally suited filters, UVIT also offers a decent angular resolution ($\sim 1.4^{\prime\prime}$) that helps to reduce the deblending issue in high-redshift objects. Here, we use deep UVIT NUV imaging of the Great Observatories Origins Survey (GOODS) Northern field combined with archival HST/UVIS and KPNO U bands photometry to estimate $\beta$ for galaxies at redshifts between 0.40 and 0.75\,. The primary aim has been to measure $\beta$ for galaxies in this least explored redshift regime and connect the cosmic evolution of $\beta$ between local and high-z galaxies. Additionally, we take advantage of UVIT's sensitivity to identify galaxies to fainter magnitudes (M$_{1500} \lesssim - 15.6$) and explore the nature of low-luminosity galaxies that contribute the bulk of the UV luminosity at any redshift \citep{kurcz2014}.

The paper is arranged as follows: we discuss the data and observations in Section \S\ref{s_data}, photometry and sample selection in \S\ref{s_selection}, analysis in \S\ref{s_analysis}, the discussion is presented in \S\ref{s_discussion}, followed by a summary in \S\ref{s_summary}. Throughout the paper, all magnitudes are in the AB system, and we adopt a cosmology with $H_0 = 70$ km s$^{-1}$ Mpc$^{-1}$, $\Omega_{\Lambda} = 0.7$, $\Omega_M = 0.3$\,.

\section{Data and Observations}
\label{s_data}
We use the UVIT NUV broadband and medium band observations of the GOODS-north deep field (i.e., AstroSat UV Deep Field North - AUDFn; \citet{mondal2023}) to sample a part of the rest-frame FUV spectra of galaxies at $z =$ 0.40 $-$ 0.75\,. The UVIT is a UV telescope on the AstroSat satellite, India's first multi-wavelength space observatory \citep{singh2014,kumar2012}. It is equipped with multiple filters in the FUV and NUV channels with the capability of simultaneous observations. More details about the instrument and its calibration can be found in \citet{tandon2017,tandon2020}. We observed the GOODS-North field using two broadband (F154W, N242W) and one medium band (N245M) filters over multiple orbits of the satellite covering UT 2018-03-10 to 2018-03-12 (proposal ID - G08\_077, PI - Kanak Saha). In this study, we use images acquired with the NUV filters (N242W, N245M) only. We utilise the CCDLAB pipeline \citep{postma2017,postma2021} to produce science-ready images from the raw UVIT data. A detailed discussion of these UVIT observations and data products is included in the AUDFn catalog paper \citep{mondal2023}. The UVIT imaging data covers a $\sim$ 616 arcmin$^2$ field, including the 157.8 arcmin$^2$ science-area observed by HST CANDELS \citep{grogin2011,koekemoer2011}, with an angular resolution $\sim$ 1\farcs4 (Figure~\ref{beta_map}). Apart from the UVIT data, we have used archival HST \textit{UVIS} F275W, F336W band (HDUV survey; \citet{beckwith2006,oesch2018}) and KPNO U band photometric catalog in this work. Details of the photometric bands are listed in Table \ref{table_bands}. The survey footprints for each of these observations in the GOODS-north field are shown in Figure~\ref{beta_map}. The HST photometric catalog, prepared by \citet{oesch2018}, is obtained from \textit{MAST}  \footnote{https://archive.stsci.edu/prepds/hduv/}. The KPNO 4m Mayall telescope U band photometry is acquired from the 3D-HST catalog provided by \citet{skelton2014}. The redshifts for our sample galaxies were adopted from the 3D-HST spectroscopic catalog \citep{momcheva2016} measured using WFC3 G141 grism.

\begin{table*}
\centering
\caption{Details of the photometric bands used for estimating the UV continuum slope $\beta$.}
\label{table_bands}
\begin{tabular}{p{1.5cm}p{2cm}p{1.5cm}p{1.2cm}p{2.8cm}p{1.5cm}p{3cm}}
\hline
Filter & Bandpass & $\lambda_{mean}$ & ZP mag & Unit conversion & 5$\sigma$ Depth & Telescope\\
 & (~\AA~) & (~\AA~) & (AB) & (erg~s$^{-1}$cm$^{-2}$\AA$^{-1}$) & (AB) & \\
 (1) & (2) & (3) & (4) & (5) & (6) & (7) \\\hline
N242W & 2000 $-$ 3050 & 2418 & 19.763 & 2.32$\times10^{-16}$ & 26.7 & AstroSat/UVIT\\
N245M & 2148 $-$ 2710 & 2447 & 18.452 & 7.57$\times10^{-16}$ & 26.5 & AstroSat/UVIT\\
F275W & 2195 $-$ 3187 & 2710 & 24.065  & 3.51$\times10^{-18}$ & 27.4 & HST WFC3/UVIS\\
F336W & 2946 $-$ 3797 & 3355 & 24.615  & 1.38$\times10^{-18}$ & 27.8 & HST WFC3/UVIS\\
U & 3000 $-$ 4500 & 3593 & 31.369 & 2.39$\times10^{-21}$ & 26.4 & KPNO 4m Mayall\\\hline
\end{tabular}

\textbf{Note.} Table columns: (1) name of the photometric filter; (2) filter bandpass in \AA; (3) filter mean wavelength in \AA; (4) filter zero point magnitude in the AB system; (5) filter unit conversion factor in CGS; (6) the value of 5$\sigma$ detection limit (Source: N242W, N245M - \citet{mondal2023}; F275W, F336W - \citet{oesch2018}; U - \citet{skelton2014}) (7) the telescope used for observation.

\end{table*}

\begin{figure}
    \centering
    \includegraphics[width=3.5in]{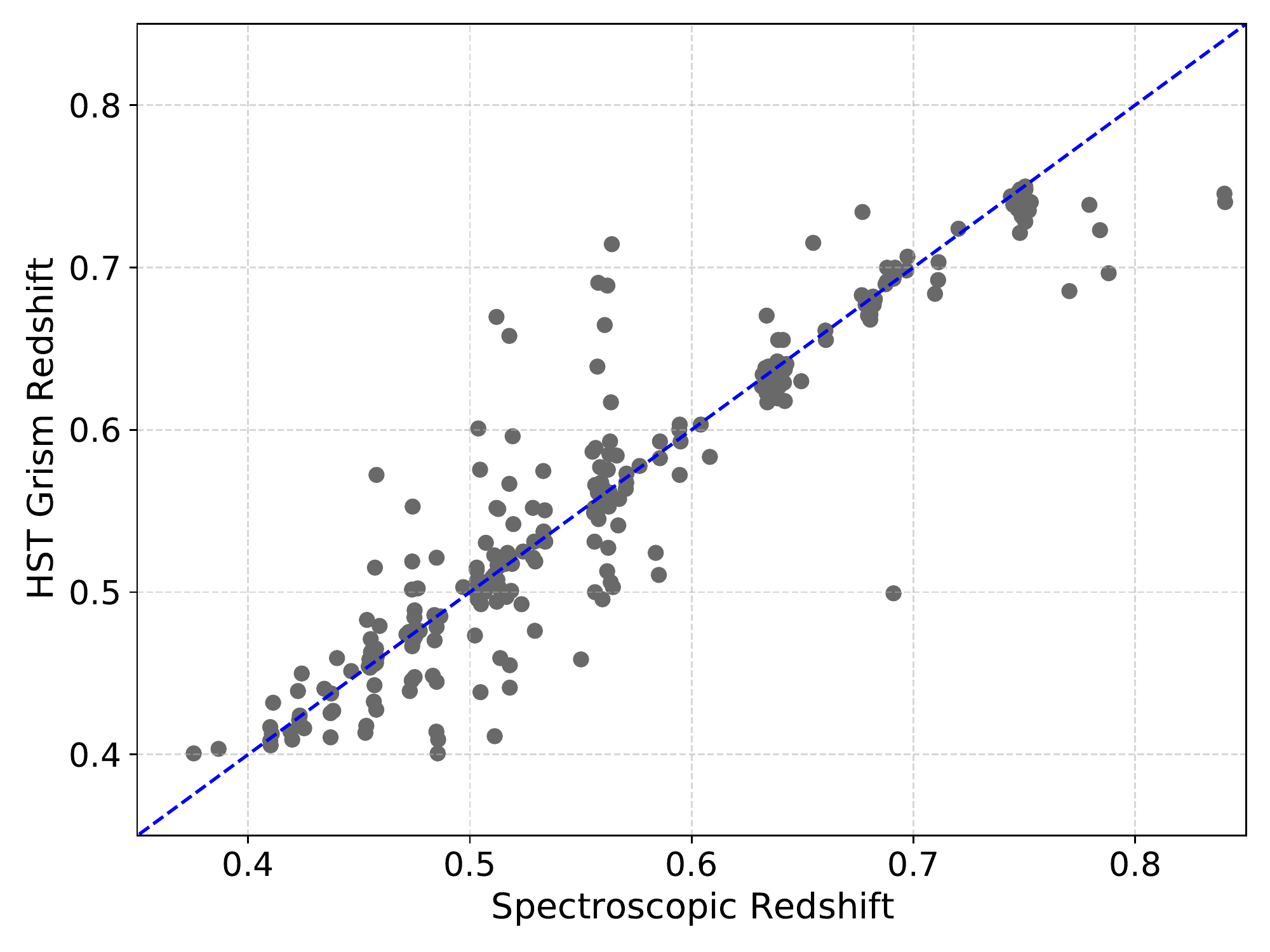} 
    \caption{The galaxies (among the selected sample) that have both reliable grism redshift (\textit{use\_grism} = 1 from 3D-HST spectroscopic catalog; \citet{momcheva2016}) and spectroscopic redshift (from \citet{skelton2014}) are shown. The blue dashed line shows the line of equality. The figure shows good agreement between the two redshift values for the majority of our sample.}
    \label{beta_redshift}
\end{figure}

\begin{figure}
    \centering
    \includegraphics[width=3.5in]{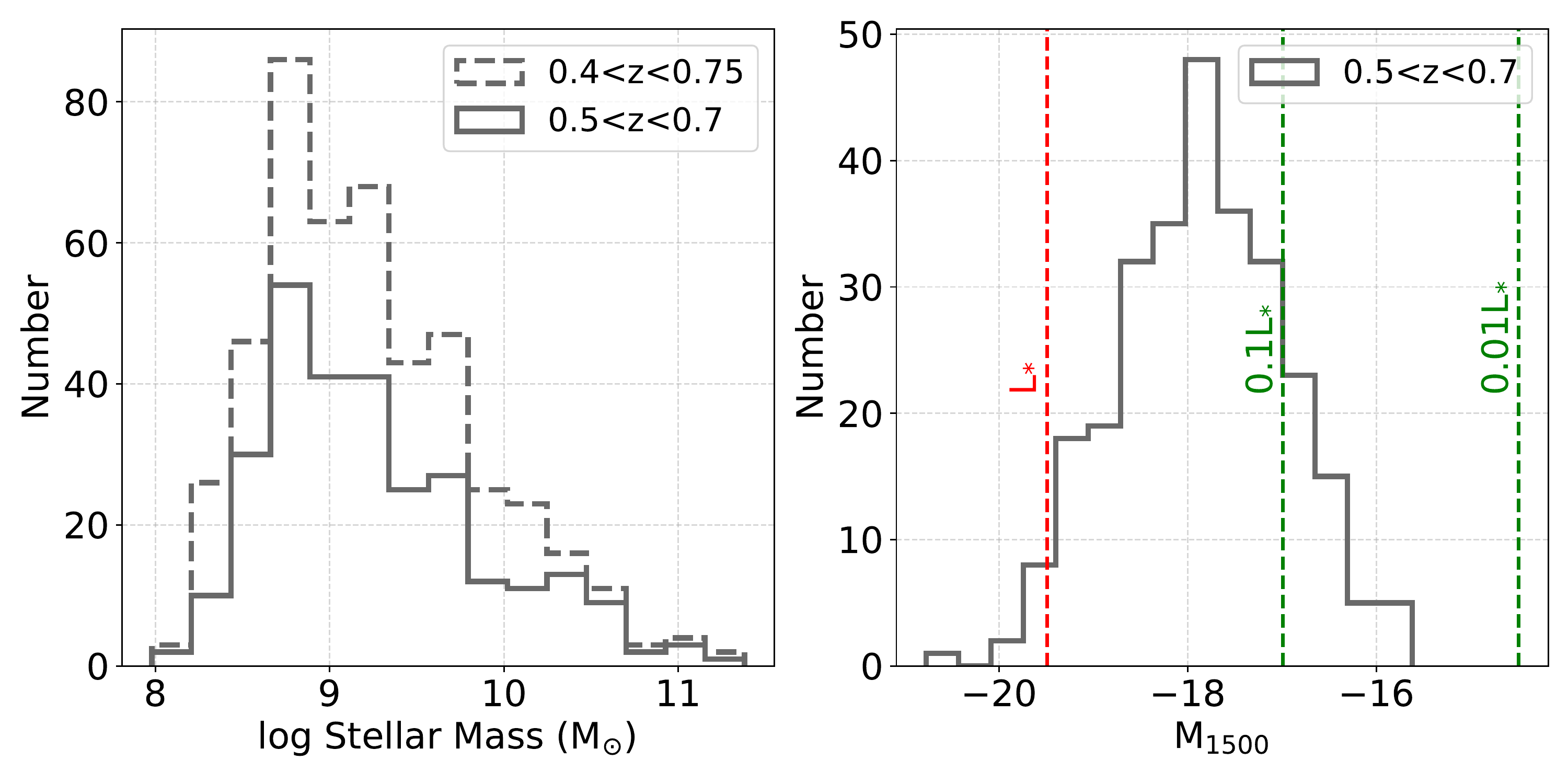} 
    \caption{Left: The stellar mass distribution of 465 sample galaxies from the 3D-HST photometric catalog \citep{skelton2014}. The histogram for the entire sample is shown in dashed line whereas the solid one represents samples within redshift range {\it z} $=$ 0.5 $-$ 0.7\,. Right: The UV absolute magnitude (M$_{1500}$) of galaxies between redshift of 0.5 and 0.7\,. M$_{1500}$ is estimated using the N242W or N245M band UVIT observations (details in \S\ref{s_muv}). The M$_{1500}$ values corresponding to $L^*$, 0.1$L^*$, and 0.01$L^*$ luminosity are marked with vertical dashed lines assuming M$^{*} = -19.49$ (for redshift range $z = 0.4 - 0.6$) from \citet{arnouts2005}.}
    \label{mass_muv}
\end{figure}

\section{UVIT photometry and Sample selection}
\label{s_selection}
The photometry on the UVIT NUV images is compiled using SExtractor \citep{bertin1996}. We use the N242W broadband image to identify sources across the field. The minimum area for detection is fixed as 10 pixels considering the FWHM of the UVIT PSF (which is $\sim$ 3.5 pixels). The initial source photometry was corrected for object blending and spurious detections are cleaned from the final catalog. A comprehensive discussion on background estimation, source detection, and catalog parameters of the GOODS-north UVIT field is included in \citet{mondal2023}. The photometry of the identified objects was performed on the background-subtracted images of the respective bands. For N245M, we used the detections from N242W and performed photometry on the N245M band image utilizing the dual mode function of SExtractor. The FWHM of the PSF in both the NUV images are found to be $\sim$1\farcs2\,. In this study, we consider magnitudes estimated with a fixed circular aperture of radius 1\farcs4 including aperture correction for the UVIT filters (from \citet{mondal2023}). The aperture correction factor is estimated using the PSF growth curve in each individual band, to account for the source flux outside the aperture. A similar approach has been adopted in \citet{pannella2015} to sample the photometry of galaxies from images with different resolutions.

To include a galaxy in our sample within the GOODS-north field, we have checked several criteria. First, we use the 3D-HST photometric catalog to separate sources that do not have any other detection within 1\farcs4 radius. We find 26011 such sources out of a total of 38279 listed in the HST catalog. This radius corresponds to the upper limit of FWHMs for a sample of stars selected across the UVIT field to construct the NUV PSF \citep{mondal2023}. This provides a list of HST sources that do not have source confusion within the UVIT PSF. From this sample, we selected sources with reliable redshift measurements (i.e., with \textit{use\_grism = 1}) within the redshift range 0.40 $-$ 0.75 using the 3D-HST spectroscopic catalog \citep{momcheva2016}. The redshift range is fixed such that either of the two UVIT NUV filters can sample the rest-frame FUV continuum of the galaxies while avoiding the Ly$\alpha$ line. This is discussed in detail in  \S\ref{s_analysis_1}. We cross-match sources between the selected HST sample and the UVIT NUV photometric catalog (from \citet{mondal2023}) using a 1\farcs4 matching-radius. We find 465 common sources and consider those as our galaxy sample. We do note here that $\sim$98\% (91\%) of the final selected galaxies are brighter than the 5$\sigma$ (10$\sigma$) detection limit in N242W, which confirms a good S/N of the chosen sample. We also use HST F275W and F336W magnitudes from the HDUV catalog using the unique HST source ID to cross-match sources with 3D-HST. The KPNO U band magnitude of the selected sources is directly taken from the 3D-HST catalog \citep{skelton2014}.

Among 465 final sample galaxies, 261 have spectroscopic redshift measurements as well from \citet{skelton2014}. In Figure~\ref{beta_redshift}, we compare both redshift values and demonstrate good agreement between them. We use the redshift measurements from the HST grism for all sources in this study. Apart from providing more samples, we prefer to use HST grism redshift as these measurements include a reliability flag unlike the spectroscopic redshift values listed in \citet{skelton2014}. We show the total stellar mass and UV absolute magnitude (M$_{1500}$) of the sample galaxies in Figure~\ref{mass_muv}. The mass of 465 selected galaxies, taken from the 3D-HST catalog \citep{skelton2014}, ranges between $\sim 10^8$ and $10^{11}$ M$_{\odot}$. The M$_{1500}$, shown for a sub-sample of galaxies between redshift of 0.50 and 0.70 (explained in \S\ref{s_muv}), has a range between $\sim -21.0$ and $-15.6$ mag. Both the figures together characterize the types of galaxies selected in this study.

\section{Analysis}
\label{s_analysis}

\subsection{Estimating UV spectral slope $\beta$}
\label{s_analysis_1}

The UV continuum flux (F$_{\lambda}$) of a galaxy is characterised by a power law of the form  F$_{\lambda}$ $\propto$ $\lambda^{\beta}$. The value of $\beta$ (i.e., the spectral slope), determines the shape of the continuum (where $\beta=-2$ refers to a flat spectrum in $f_{\nu}$). Slope $\beta$ is measured within a wavelength range that covers both the FUV and NUV waveband in the rest-frame. \citet{calzetti1994} defined the wavelength range as 1268 $-$ 2580~\AA~ which has also been followed by others to sample the UV spectral slope. We targeted a redshift range such that the photometric filters effectively sample the spectral energy distribution within 1215 $-$ 2600~\AA~ (the same has been considered in \citet{reddy2018}). At z=0.4, the blue end and the effective wavelength of the N242W filter fall at 1428~\AA~ and 1727~\AA~, respectively. We have not considered galaxies below redshift 0.4, where N242W samples redward of the FUV part of the spectra. At {\it z} = 0.4, the effective wavelength of the reddest filter (i.e., U) corresponds to 2566~\AA~, satisfying the defined range to sample $\beta$. The blue end of the N242W filter crosses 1215~\AA~ (Ly$\alpha$ line) at $z =$ 0.65\,. To avoid the possible contribution of Ly$\alpha$ emission, we have used the N242W band measurements up to redshift 0.64 and further considered the medium band N245M filter in the range 0.64~$<$~z~$<~$0.75\,. We fix the upper limit of redshift to 0.75, beyond which Ly$\alpha$ would appear in the N245M filter. Therefore, we consider two sets of filters for measuring $\beta$ of 465 sample galaxies in two redshift ranges. For redshift between 0.40 $-$ 0.64, the filters N242W, F275W, F336W, and U are employed, while we use N245M, F275W, F336W, and U for the range 0.64 $-$ 0.75\,. 

\begin{figure}
    \centering
    \includegraphics[width=3.5in]{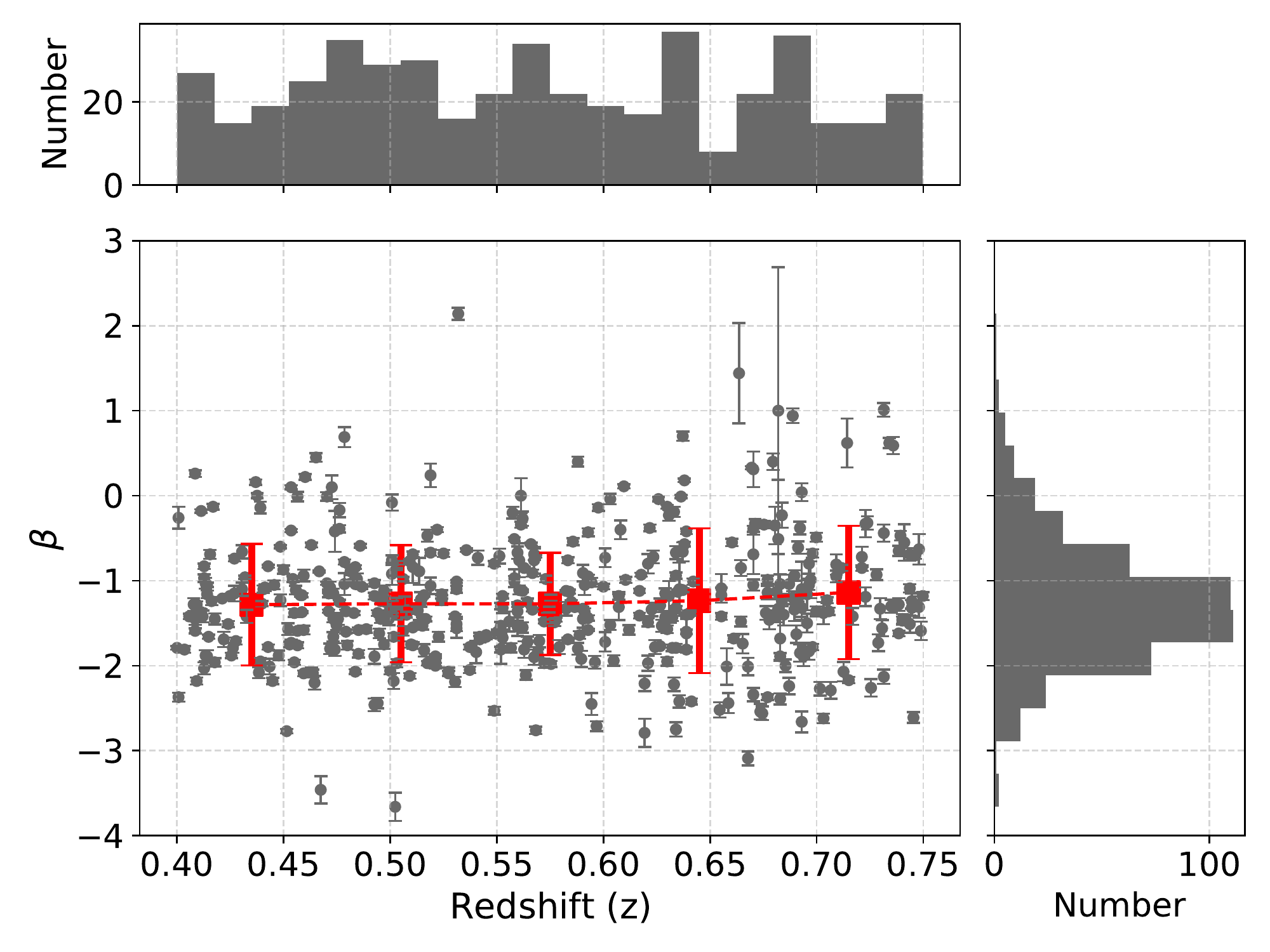} 
    \caption{Measured UV spectral slope ($\beta$) of the sample galaxies (grey points) versus redshift. The error bar associated with each grey point indicates the uncertainty in $\beta$ due to the photometric error in observed fluxes. The red points show the sample mean of $\beta$ calculated in five redshift bins as defined in \S\ref{s_analysis_2}. The associated error bar shows the 1$\sigma$ scatter in each respective bin. The histogram of redshift and $\beta$ for the selected samples are displayed in the adjacent panels. Around 98\% of the total sample plotted here is brighter than the 5$\sigma$ detection limit in N242W.}
    \label{beta_value}
\end{figure}

As the HST F275W and F336W bands cover a smaller field (Figure~\ref{beta_map}), there are 249 galaxies in our sample which have photometry available in only three bands (N242W, N245M, U). The remaining sample ($\sim$46\% of the total) has measurements in four or more filters including HST. In the case of galaxies not covered by HST F275W and F336W observations, we only have two photometric points (either N242W or N245M, and U) to evaluate $\beta$. The magnitudes in each band are first corrected for the Galactic extinction in the observed frame. The extinction in V band (A$_{V}$) along the line of sight to each galaxy is obtained from the Schlegel \citep{schlegel1998} Galactic extinction map using the query interface of the \textit{NASA/IPAC Infrared Science Archive} \footnote{https://irsa.ipac.caltech.edu/applications/DUST/}. We adopt the Fitzpatrick \citep{fitzpatrick1999} Galactic extinction law with $R_{V}$ = 3.1 to estimate extinction at the effective wavelength of each filter from A$_{V}$. The extinction corrected magnitudes are then converted to flux (F$_{\lambda}$ in erg~s$^{-1}$cm$^{-2}$\AA$^{-1}$) using the zero point magnitude and unit conversion factor listed in Table \ref{table_bands}. After shifting the wavelengths to the galaxy restframe, given the redshift, we plot the photometric measurements for each galaxy in the logF$_{\lambda}$ - log$\lambda$ plane. We perform a least square linear fit to these data points and the slope of the fitted line is the UV spectral slope, $\beta$ \citep[see,][]{jiang2020,reddy2018}. In Figure~\ref{beta_value}, we show the measured $\beta$ and the redshift of the sample galaxies. The majority of the galaxy has $\beta$ ranging between $-1$ to $-2$. We followed the error propagation method to estimate the error in $\beta$ (as shown in Figure~\ref{beta_value}) from the photometric error of observed flux values used in the fitting.

\begin{table}
%\centering
\caption{The estimated value of $\langle\beta\rangle$ and 1$\sigma$ scatter in different redshift bins.}
\label{table_beta_dist}
\begin{tabular}{p{2cm}p{1.5cm}p{2.0cm}p{1cm}}
\hline
Redshift & Sample & \centering $\langle\beta\rangle$ & 1$\sigma$\\
range & (N) &  & scatter\\
(1) & (2) & \centering(3) & (4)\\\hline
0.40 $-$ 0.47 & 86 & $-1.40\pm0.05$ & 0.63\\
0.47 $-$ 0.54 & 110 & $-1.36\pm0.05$ & 0.52\\
0.54 $-$ 0.61 & 97 & $-1.37\pm0.06$ & 0.55\\
0.61 $-$ 0.68 & 84 & $-1.32\pm0.08$ & 0.76\\
0.68 $-$ 0.75 & 88 & $-1.23\pm0.12$ & 0.64\\\hline
0.40 $-$ 0.75 & 465 & $-1.33\pm0.07$ & 0.60\\\hline
\end{tabular}
\textbf{Note.} Table columns: (1) the redshift range; (2) number of sample galaxies within the redshift window; (3) mean value of $\beta$ in the redshift bin; (4) the related 1$\sigma$ scatter in $\beta$ for galaxies sampled within the redshift bin.  
\end{table}

\begin{figure*}
    \centering
    \includegraphics[width=7in]{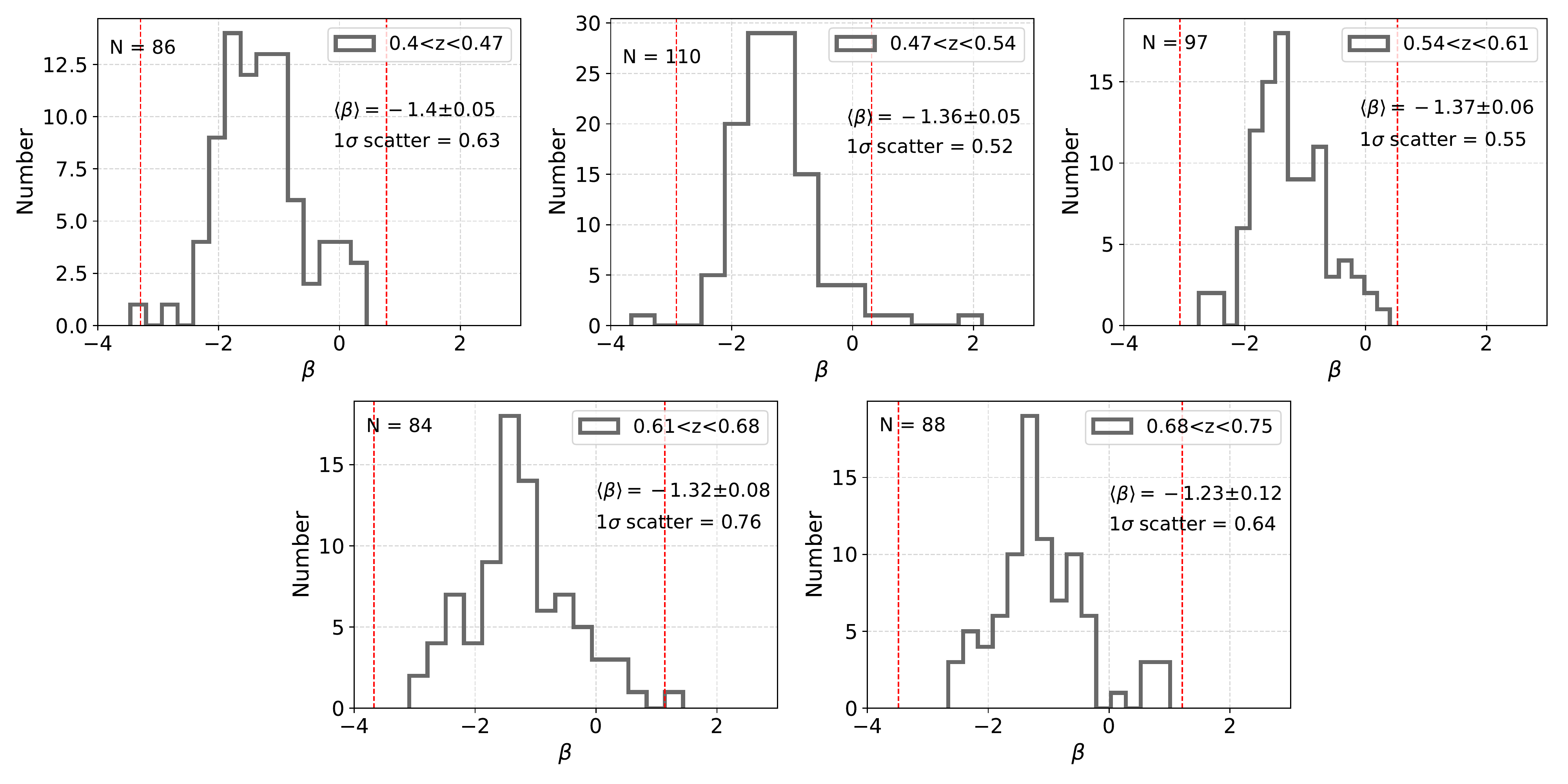} 
    \caption{The distribution of the observed $\beta$ for sample galaxies within five different redshift bins (as defined in \S\ref{s_analysis_2} and shown in Figure~\ref{beta_value}). The total number of samples (N), mean $\beta$ with error, and the 1$\sigma$ scatter in $\beta$ (from sigma-clipped gaussian fitting) for each bin are noted (Table \ref{table_beta_dist}). The vertical red dashed lines mark the limit of 3-sigma iterative clipping on either side of the mean. The mean and scatter are estimated for points between the red lines.}
    \label{beta_dist}
\end{figure*}

\subsection{Evolution of $\beta$ with Redshift}
\label{s_analysis_2}

\begin{figure}
    \centering
    \includegraphics[width=3.5in]{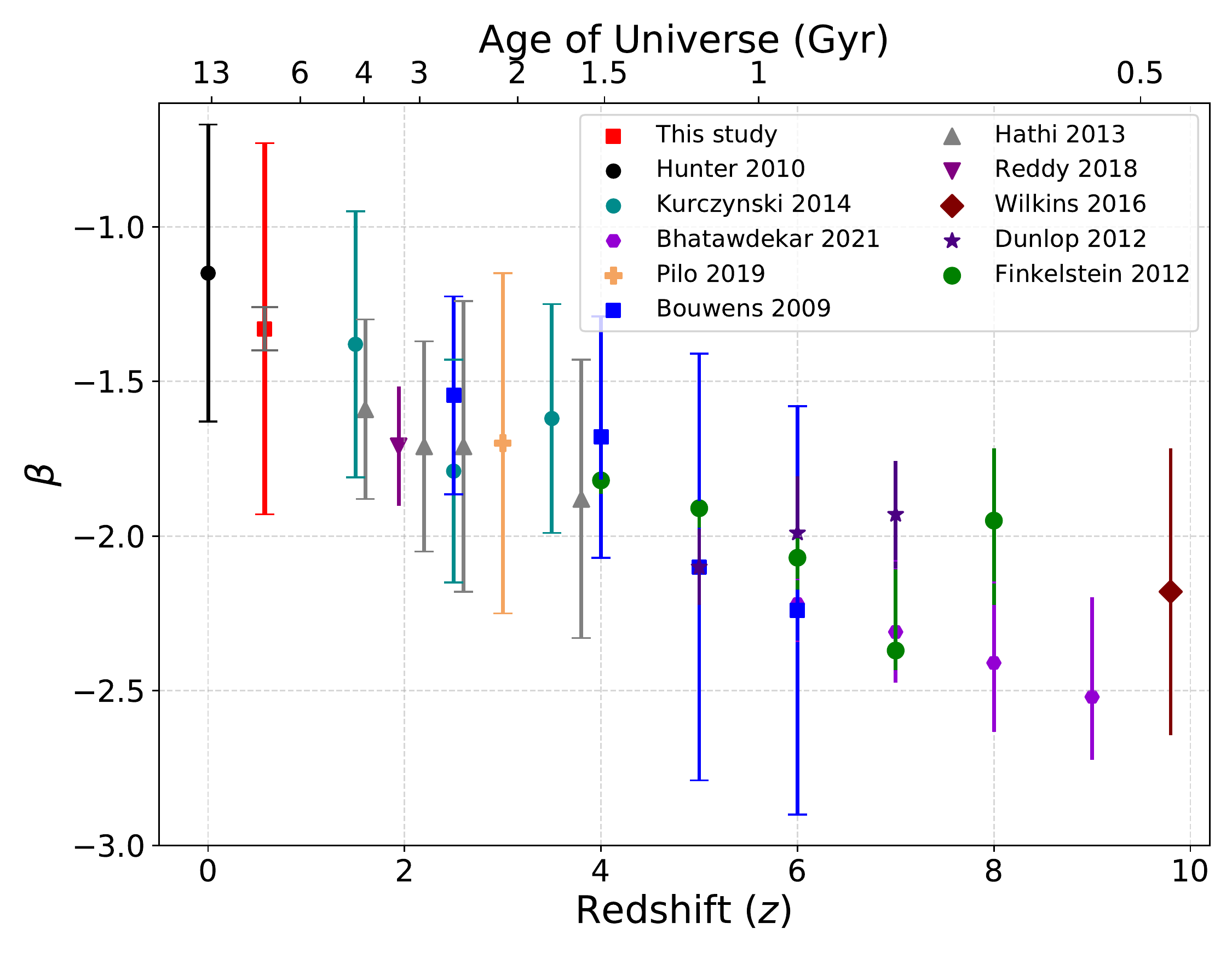}
    \caption{Evolution of $\beta$ with redshift. The mean value of $\beta$ estimated in the 0.40 $-$ 0.75 redshift range is shown by the red point. The grey and the red error bars associated with the red point indicate photometric error and 1$\sigma$ scatter in $\beta$, respectively. The mean/median value of $\beta$ reported in other studies at different redshifts is also shown. The 1$\sigma$ scatter in $\beta$ is shown as error bars with a cap for \citet{hunter2010}, \citet{kurcz2014}, \citet{pilo2019}, \citet{bouwens2009}, and \citet{hathi2010}. In case the scatter was not reported, we show the associated uncertainty in $\beta$ as error bars without a cap (for \citet{bhatawdekar2021}, \citet{wilkins2013}, \citet{dunlop2013}, and \citet{finkelstein2012}). The age of the universe (in Gyr), estimated using FlatLambdaCDM subclass of astropy.cosmology package, is shown on the top axis.}
    \label{beta_ref}
\end{figure}

To assess any evolution of $\beta$ with redshift within our sample, we consider five equal-width bins (with $\Delta z = 0.07$) that span the 0.40 $-$ 0.75 redshift range. We estimate the sample mean and 1$\sigma$ scatter for each bin, shown in Figure~\ref{beta_value} as red points. Neither the mean nor scatter show much variation within the redshift range considered. In Figure~\ref{beta_dist}, we show the histograms of $\beta$ for galaxies falling in each redshift bin. We fit Gaussian functions to each histogram, perform iterative 3-sigma clipping to get rid of outliers, and then estimate mean and standard deviation (1$\sigma$ scatter) of $\beta$ in each bin (Figure~\ref{beta_dist}). We also calculate the mean error in $\beta$ from the photometric uncertainties for each bin. In Table~\ref{table_beta_dist}, we list the number of galaxies (N), $\langle\beta\rangle$ with error, and 1$\sigma$ scatter for each of the five redshift bins. The mean $\beta$ stays nearly constant at $\sim-1.3$, with no evidence for any significant evolution of the UV spectral slope within the considered redshift range. There are 54 galaxies that show extreme $\beta$ values ($\beta<-$2), which can be interesting to study in detail. Galaxies with such extreme $\beta$ values are more common at higher redshifts. This is possible for systems with a very young stellar population, minimum dust, and lower metallicity. Therefore, $\beta < -2$ in our sample could imply such unique galaxy populations which can still be present at a lower redshift.

In order to compare the estimated $\beta$ value with measurements at different redshifts, we calculate the average $\beta$ for the full 0.40 $-$ 0.75 redshift range. Following a similar approach, we find mean, median and 1$\sigma$ scatter of $\beta$ as $-1.33\pm0.07, -1.32$ and 0.60 respectively (Table \ref{table_beta_dist}). In Figure~\ref{beta_ref}, we show our measurements along with values of $\beta$ reported across the redshift range of $\sim$ 0 $-$ 10\,. The measurements from other studies in Figure~\ref{beta_ref} cover galaxies of different luminosities. The majority of the studies explored UV spectral slopes of galaxies at redshifts higher than 1 using HST (discussed in Section \ref{s_intro}), whereas the UV spectral slope of a sample of local dwarf galaxies \citep{hunter2010} has been estimated using their GALEX (FUV$-$NUV) observed color by \citet{kurcz2014}. 

Our results shown in Figure~\ref{beta_ref} supports the progressive evolution of observed $\beta$ with redshift. The UV continuum of galaxies has evolved towards redder slopes with cosmic time, signifying gradual dust enrichment in galaxies. Our mean $\beta$ is found to be bluer compared to the value reported for local dwarf galaxies (i.e., $\langle\beta\rangle =-1.15$, \citet{kurcz2014,hunter2010}). This further highlights the importance of our measurement to connect the cosmic evolution of $\beta$ by filling the $\sim$ 8 Gyr gap between {\it z} $\sim$ 1 and {\it z} $\sim$ 0\,. As the value of $\beta$ can also depend on the luminosity of a galaxy, the trend of mean $\beta$ with redshift shown in Figure \ref{beta_ref} can be biased depending on the diversity of luminosity range studied by different authors. To better constrain the $\beta - z$ relation, one must consider a specific luminosity range and estimate mean $\beta$ from galaxies belonging to that range. In Section \S\ref{s_muv}, we have considered the luminosity range $-20~<~$M$_{1500}~<~-19$ and shown the $\beta - z$ evolution in Figure \ref{beta_abs_specific}.

We notice a larger scatter in $\beta$ around the estimated mean. This may indicate the diversity in dust properties and/or star formation history in galaxies of different luminosities within the considered redshift range. We do note here that 216 galaxies out of the total selected sample have photometry available in four or more passbands. Considering only these galaxies, we find a mean (median) and 1$\sigma$ scatter in $\beta$ of $-1.42\pm0.08~(-1.39)$ and 0.55 respectively between redshift of 0.40 and 0.75\,. The similar value of the computed mean $\beta$ signifies that our measurements with only two photometric points are robust. The decrease in scatter does mean, however, that additional photometric data points could result in fewer outliers in the $\beta$ measurements.

\subsection{Correlation of $\beta$ with M$_{1500}$}
\label{s_muv}

\begin{figure}
    \centering
    \includegraphics[width=3.5in]{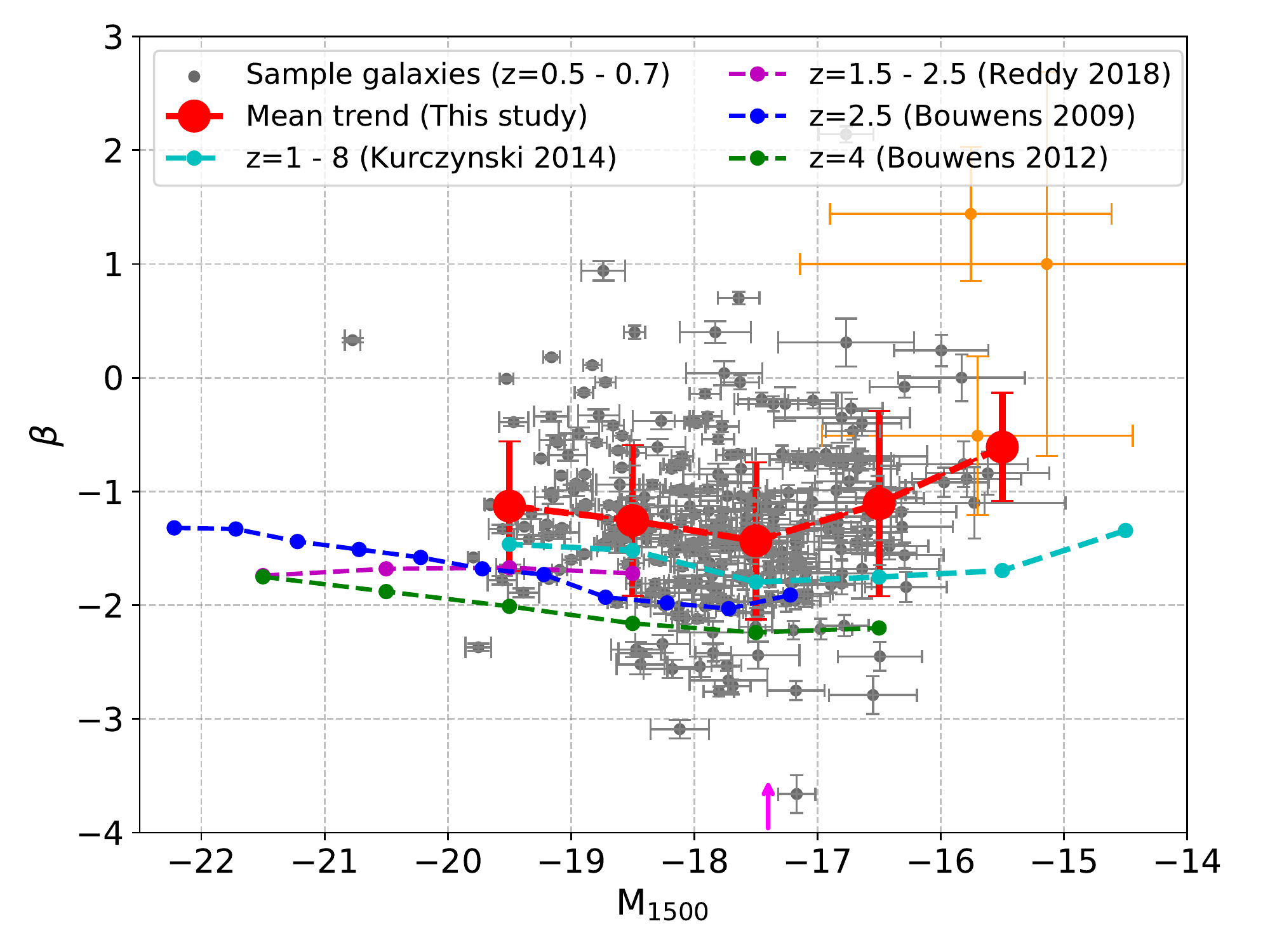}
    \caption{UV slope $\beta$ versus UV luminosity (absolute magnitude). The grey dots show the UV continuum slope $\beta$ and the rest-frame UV absolute magnitude (M$_{1500}$) of sample galaxies between redshift of 0.50 and 0.70\,. The red filled circles indicate the mean $\beta$ within the defined M$_{1500}$ bins as listed in Table \ref{table_muv}. The associated error bars signify 1$\sigma$ scatter in $\beta$. For the faintest luminosity bin (i.e., $-16 <$ M$_{1500} < -15$), the plotted mean value is estimated using 7 samples (excluding the three orange points with large error) belonging to that bin. The cyan, magenta, blue, and green dashed lines show the reported $\beta$ - M$_{1500}$ values from \citet{kurcz2014}, \citet{reddy2018}, \citet{bouwens2009}, \citet{bouwens2012} respectively. The absolute magnitude reported by \citet{kurcz2014} is measured as M$_{2330}$ which on average they found as 0.1 mag brighter than M$_{1500}$. The magenta arrow marks the faint limit of detection in GOODS-N and GOODS-S field by \citet{reddy2018} between redshift of 1.5 and 2.5\,.}
    \label{beta_abs}
\end{figure}

We estimated the rest-frame UV absolute magnitude of a sub-set of the selected samples by carefully choosing a redshift range $z = 0.5 - 0.7$ such that the effective wavelengths of the UVIT filters (N242W and N245M) fall in the rest-frame 1400 $-$ 1600~\AA~ range. In Figure~\ref{beta_abs}, we show the distribution of observed $\beta$ and M$_{1500}$ for 282 sample galaxies falling within the selected redshift window. We detect galaxies with M$_{1500}$ spanning wide range of magnitudes from $\sim$ $-$20.8 to $-$15.6 mag (i.e., $\sim$ 3.3~$L^{*}_{z=0.5}$ $-$ 0.028~$L^{*}_{z=0.5}$; as per the M$_{*}$ provided in \citet{arnouts2005} at $z = 0.4 - 0.6$). On the fainter side, we detect galaxies down to $\sim$0.028~$L^*$, which highlights the superior sensitivity of UVIT in detecting fainter galaxies. The galaxies that are more luminous in UV (M$_{1500}<-$19 mag) have observed $\beta>-2$, whereas extreme $\beta$ values ($\beta<-2$) are mostly associated with lower luminosity galaxies with $-$17~$<~$M$_{1500}~<-$19 mag. We consider five M$_{1500}$ bins between $-$20.0 and $-$15.0 mag and plot the sample mean and 1$\sigma$ scatter for each bin in Figure~\ref{beta_abs} (red points). The mean value of UV spectral slope shows a slightly decreasing trend with increasing M$_{1500}$ between $-$21 and $\sim-$18 mag (spearman coefficient = $-$0.26), whereas on the fainter side, i.e., fainter than M$_{1500}=-$18 mag, the trend in  mean $\beta$ reverses (spearman coefficient = 0.21). We note here that the 3 samples shown in orange (in the faintest bin) are excluded while measuring the quantities mentioned above due to their large error in M$_{1500}$.

\begin{figure*}
    \centering
    \includegraphics[width=6.5in]{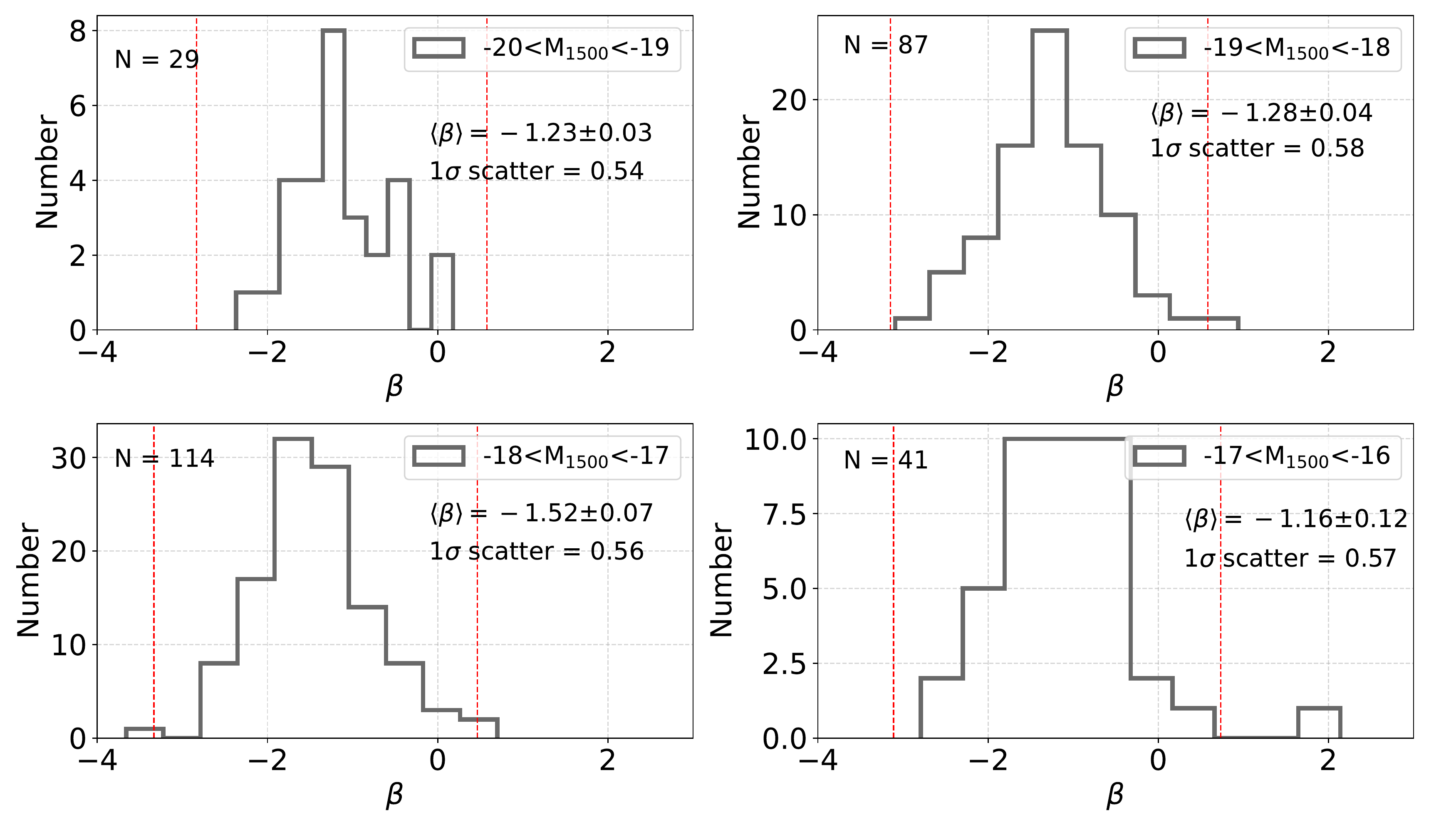} 
    \caption{The distribution of $\beta$ for galaxies within four different M$_{1500}$ bins between redshift of 0.50 and 0.70\,. The total number of samples (N), mean $\beta$ with error, and the 1$\sigma$ scatter in $\beta$ (from sigma-clipped gaussian fitting) for each luminosity bin are noted (Table \ref{table_muv}). The vertical red dashed lines mark the limit of 3-sigma iterative clipping on either side of the mean.}
    \label{beta_abs_dist}
\end{figure*}

To estimate the mean and scatter avoiding sample outliers for the M$_{1500}$ distributions of the first four bins as shown in Figure~\ref{beta_abs_dist}, we followed a similar approach as in \S\ref{s_analysis_2}. The number of galaxies, mean $\beta$ and 1$\sigma$ scatter for each bin are listed in Table \ref{table_muv}. We do not show the bin $-16<$ M$_{1500}<-15$, as it contains only 7 galaxies. We notice no strong consistent relation between $\beta$ and M$_{UV}$ for our entire luminosity range. Our UVIT observation in the GOODS-north field has detected faint galaxies down to M$_{1500} = -15.6$ at redshift 0.5 $-$ 0.7\,. In the HDUV survey of GOODS-S and GOODS-N fields, \citet{reddy2018} detected galaxies down to M$_{1500} = -17.4$ mag (magenta arrow in Figure~\ref{beta_abs}), on the fainter side, for the redshift range $z = $ 1.5 $-$ 2.5\,. The faintest galaxies (M$_{1500} > -16$), detected in our study, show relatively redder $\beta$ slope. We do not see extreme blue $\beta$ values for the faintest galaxies either.

\begin{figure}
    \centering
    \includegraphics[width=3.5in]{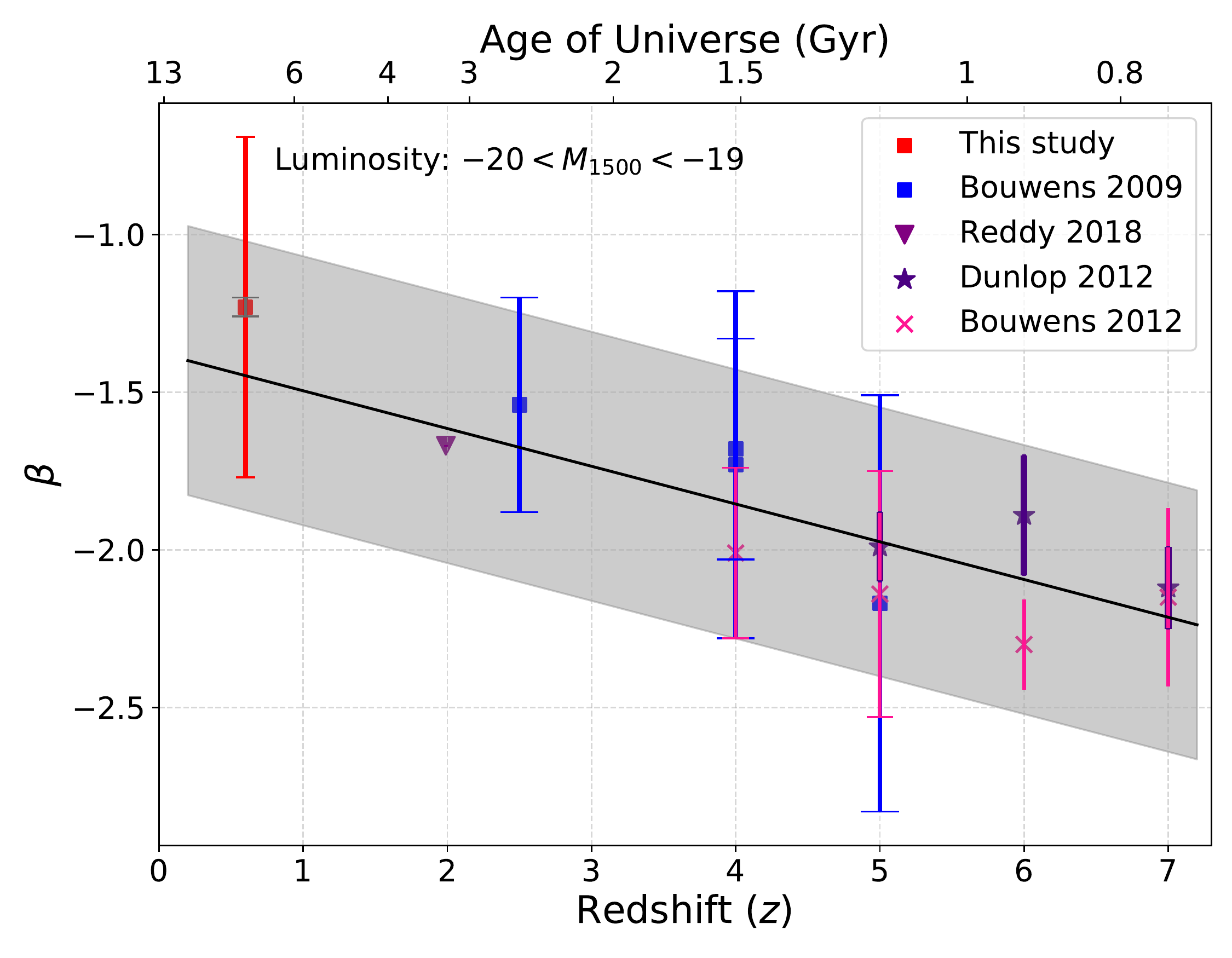}
    \caption{A plot similar to Figure~\ref{beta_ref} is shown for $\beta$ measured within a particular luminosity range $-20<$M$_{1500}<-19$\,. The black line shows the least square linear fit to the points acquired from other studies. The grey-shaded region around the black line shows mean 1$\sigma$ scatter as computed from these studies. The red point shows the mean $\beta$ measured in our study within the specific luminosity range (from Table \ref{table_muv}). The red and the grey error bars associated with the red point denote 1$\sigma$ scatter and photometric error in $\beta$ respectively. The age of the universe (in Gyr), estimated using FlatLambdaCDM subclass of astropy.cosmology package, is shown on the top axis.}
    \label{beta_abs_specific}
\end{figure}

The measurements shown in Figure~\ref{beta_ref} include galaxies of different luminosity. To confirm our inference on the progressive evolution of $\beta$ with redshift for a specific luminosity range, we have considered all the galaxies with $-20 < M_{1500} < -19$~mag in our sample and plotted the mean $\beta$ in Figure~\ref{beta_abs_specific}. We also show measurements at different redshifts for the same luminosity range from other studies. We perform a least square linear fit (black line in Figure~\ref{beta_abs_specific}) to the values from the literature and extrapolate it to the redshift that corresponds to our measurement (red point). The fitted line shows a gradually increasing value of $\beta$ with decreasing redshift. The mean $\beta$ (red point), measured between $-20 < M_{1500} < -19$~mag in our study, supports the predicted trend with a slightly higher value. Our measurement falls within the limit of mean 1$\sigma$ scatter from the fitted line (i.e., inside the grey-shaded region) as reported by the studies listed in Figure \ref{beta_abs_specific}. Hence, our study provides evidence for the progressive evolution of $\beta$ with redshift for galaxies within a specific luminosity range, which further strengthens the idea of gradual dust enhancement in galaxies with cosmic time.

\begin{table}
\caption{The estimated value of $\langle\beta\rangle$ and 1$\sigma$ scatter in different luminosity (M$_{1500}$) bins.}
\label{table_muv}
\begin{tabular}{p{3cm}p{0.9cm}p{2cm}p{0.9cm}}
\hline
Luminosity & Sample & \centering$\langle\beta\rangle$ & 1$\sigma$ \\
range & (N) &  & scatter \\
(1) & (2) & \centering(3) & (4)\\\hline
$-20<M_{1500}<-19$	&	29	&	$-$1.23$\pm$0.03 &	0.54\\
$-19<M_{1500}<-18$	&	87	&	$-$1.28$\pm$0.04 & 0.58\\
$-18<M_{1500}<-17$	&	114	&	$-$1.52$\pm$0.07 &	0.56\\
$-17<M_{1500}<-16$	&	41	&	$-$1.16$\pm$0.12 &	0.57\\\hline
\end{tabular}
\textbf{Note.} Table columns: (1) the redshift range; (2) number of sample galaxies within the luminosity range; (3) mean value of $\beta$ in the luminosity bin; (4) the related 1$\sigma$ scatter in $\beta$ for galaxies sampled within the luminosity range.
\end{table}

\section{Discussion}
\label{s_discussion}

Extinction at optical wavelength can be probed using the ratio of nebular Balmer lines (H$\alpha$/H$\beta$), whereas it is not the same in UV due to the absence of such lines. Also, the value of extinction estimated from the optical nebular lines can be higher as it probes regions that host young stellar populations highly obscured by dust. On contrary, the UV continuum characterises the effect of dust on the bulk of the UV emission contributed by slightly evolved young stars in galaxies \citep{calzetti1994}. As a result, the UV continuum slope remains the best proxy to determine the dust properties of galaxies, particularly in the higher redshift, where H$\beta$ line measurement becomes progressively difficult. Measuring $\beta$ at different redshift can probe the cosmic evolution of dust as galaxies continue to evolve and transform.

Our results on the UV continuum slope of galaxies highlight two key points. First, we estimate $\beta$ for a large sample of galaxies between a redshift range 0.40 $-$ 0.75, which is least explored in the context of cosmic $\beta - z$ relation. Hence, our measurements make a valuable inclusion to the existing values of observed $\beta$ at different redshifts. The redshift window is crucial, also because it forms a bridge between $z =$ 2, where the cosmic SFRD peaks \citep{zavala2021}, and the local universe. The results infer the nature of dust and star formation in galaxies at an epoch where SFRD shows a declining trend \citep{pillepich2018,zavala2021}. The mean $\beta$ estimated from our sample supports the trend of global $\beta - z$ relation as reported in the literature (Figure~\ref{beta_ref}~\&~\ref{beta_abs_specific}). Our study further highlights the unique advantage of UVIT NUV imaging in deciphering FUV characteristics of galaxies at redshift $\sim$ 0.5\,. As the F225W UVIS filter has not been used in the HDUV survey of the GOODS-north field, UVIT NUV observation provides the best-suited data to probe rest-frame FUV emission of galaxies at this particular redshift range. Second, utilizing the sensitivity of UVIT, we could detect galaxies faint up to 0.028 $L^*$ (corresponding to M$_{1500}$ = $-$15.6 mag) and estimate their $\beta$. Exploring extinction in fainter galaxies is crucial because low-luminous galaxies are abundant in number and they play a major role in fulfilling the total cosmic UV photon budget. The fact that we report a redder slope at lower luminosity signifies that the low-mass galaxies at this redshift either contain more dust or they are less star-forming. The HDUV survey of the GOODS-north field could reach down to M$_{UV} = -17.4$ mag on the fainter side for a redshift range $z = 1.5 - 2.5$ \citep{reddy2018}, whereas \citet{kurcz2014} detected galaxies down to M$_{UV} = -14.0$ mag in the HUDF field between redshift of 1 and 2\,. Hence, our limit on the fainter side (M$_{1500} = -15.6$ mag, which is intermediate to these two surveys) shows that UVIT imaging can efficiently probe $\beta$ slope in fainter galaxies in the deep fields.  

As our study is based on samples selected with specific criteria, there can be bias involved in the results. Firstly, we use HST catalog to select objects that do not have any neighbour within a radius of 1\farcs4\,. This will clearly exclude some objects that failed to fulfil the condition within the observed field. Secondly, among the clean HST sources, we select galaxies that have reliable grism redshift. This has again excluded some objects that either do not have redshift measurement or the redshift was flagged as non-reliable. Both these criteria result in an incompleteness in the sample which includes a bias in the estimated mean and scatter of $\beta$. But as we finally have 465 galaxies, we expect the selection bias to be less dominant in our results. The first criterion to produce clean sample can exclude a source irrespective of its brightness but the second one will more preferably reject fainter galaxies (as it is more difficult to have spectroscopic redshift for fainter source). Therefore the selection bias is stronger in the fainter magnitude side.

Our sample selection also depends on the detection limit of the photometric bands that we have used to evaluate $\beta$ and this can also include bias in the derived $\beta$ - {\it z} trend in Figure \ref{beta_value}. Here, we explore the distribution of observed apparent magnitude and color of the selected sources to understand the sample incompleteness. Considering the UVIT NUV (N242W and N245M) and KPNO U bands (as these bands cover our entire sample), we found that the selected samples (except a few outliers) have observed NUV or U magnitude brighter than the respective 5$\sigma$ detection limit (listed in Table \ref{table_bands}) and (NUV $-$ U) color bluer than $\sim$ 1.0\,. Also, within this range of color and apparent magnitude, the faint sources that have a bluer color are sampled more than those with a redder color. The faintest sources with much redder color are actually not found due to the limit of photometric detection. To understand the impact of such selection on the $\beta$ - {\it z} trend shown in Figure \ref{beta_value}, we checked the apparent magnitude of the selected samples with redshift. We noticed no significant trend in U magnitude with redshift, although the mean NUV magnitude of the samples in the last redshift bin is relatively fainter. But that has not resulted in sampling only significantly bluer sources in the last redshift bin. Therefore, the photometric detection incompleteness in $\beta$ sampling does not contribute any strong bias in the  $\beta$ - {\it z} trend presented here, although, we have missed samples on the fainter and redder side due to the detection limit (as noted in Table \ref{table_bands}).

We notice a large scatter in the estimated $\beta$ value of galaxies across redshift 0.40 $-$ 0.75\,. The 1$\sigma$ scatter, measured in five different redshift bins, has a range between 0.52 $-$ 0.76 (Table \ref{table_beta_dist}, Figure~\ref{beta_dist}). Earlier studies on high redshift galaxies have also reported large scatter in the observed $\beta$ values \citep{reddy2018,rogers2014,kurcz2014,dunlop2013,castellano2012,bouwens2009}. The photometric error of the sample galaxies is one among the factors that can result in large scatter in $\beta$. The scatter may also signify wide dispersion of the dust property among the sample galaxies. Apart from dust, the UV spectral slope of galaxies can also be affected by the galaxy mass, age, and star formation history. Using VANDELS survey data of more than 500 galaxies between redshift of 2 and 5, \citet{calabro2021} found a relation between galaxy metallicity and $\beta$. The metal-rich galaxies have redder spectral slope, whereas the bluer samples are relatively metal-poor. The observed scatter in our $\beta$ measurements can also be due to the diverse metallicity in our sample galaxies.

We do not notice any strong specific relation between $\beta$ and M$_{1500}$ in Figure~\ref{beta_abs}. Such lack of correlation has also been reported by \citet{dunlop2013,finkelstein2012} at different redshift ranges, whereas \citet{bouwens2014} have found a specific trend which shows progressively bluer $\beta$ towards fainter galaxies within the range $-22< $M$_{UV}<-16$ mag. The extreme blue slope of less-luminous galaxies reported in literature hinted at the presence of an extremely blue metal-poor stellar population with minimum dust content. We do notice that the majority of the luminous galaxies (in the brightest bin $-20~<$~M$_{1500}~<~-$19) have a relatively redder slope, suggesting more efficient production of dust, whereas a good percentage of less-luminous galaxies (M$_{1500}>-17$ mag) are again redder. The redder spectra of luminous galaxies may signify a scenario where these systems have accreted metal-rich gas during the evolution. In the case of less luminous galaxies (which are plausibly low-mass systems), the stellar feedback is expected to throw out dust from the system more efficiently, resulting in lower extinction and steeper continuum. Therefore, the observed redder slope of least-luminous galaxies is most likely due to their less-active ongoing star-forming mode during this cosmic time interval.

\section{Summary}
\label{s_summary}

We have used deep UVIT NUV observations combined with archival photometry in HST F275W, F336W, and KPNO U bands of the GOODS-north deep field to study rest-frame UV spectral slope of galaxies between redshift of 0.40 and 0.75\,. The main results of this study are summarised below.
\begin{itemize}
    \item Using the UVIT source catalog, we identified 465 candidate galaxies (with 98\% sample brighter than 5$\sigma$ detection limit in N242W) in the GOODS-north field that do not have source confusion within UVIT PSF and met other criteria to be considered for measuring UV continuum slope.
    
    \item Our study supports the global trend seen in the $\beta - z$ relation by adding a data point at a unique redshift range i.e., $z=0.40 - 0.75$ that has not been explored much earlier.
    
    \item We find the mean $\beta$ to be $-1.33\pm0.07$ for the redshift range 0.40 $-$ 0.75. The 1$\sigma$ scatter in $\beta$ is estimated as 0.60\,.
    
    \item The mean $\beta$ of our sample galaxies remains almost the same showing no significant redshift evolution within the range 0.40 $-$ 0.75\,. Our measurement, with respect to the earlier studies at different redshifts, supports the cosmic evolution of $\beta$ which shows $\beta$ becoming redder with decreasing redshift.
    
    \item Our observation detect galaxies with a wide luminosity range from $\sim$ 3.3 $L^{*}_{z=0.5}$ to 0.028 $L^{*}_{z=0.5}$. The limit on the fainter side i.e., M$_{1500} \sim -15.6$ mag highlights the better sensitivity of UVIT to study fainter galaxies.
    
    \item We do not find a strong consistent trend between $\beta - $ M$_{1500}$ across the entire luminosity range. Although, we notice the majority of the most luminous galaxies (M$_{1500}<-19$ mag) to have a relatively redder slope with less scatter.
    
    \item The faintest galaxies (M$_{1500}>-16$ mag) identified in our study show a relatively redder UV continuum, which plausibly signifies their quiescent star-forming phase.
    
\end{itemize}

\acknowledgements

This work is primarily based on
observations taken by AstroSat/UVIT. UVIT project is a result of collaboration between IIA, Bengaluru, IUCAA, Pune, TIFR, Mumbai, several centres of ISRO, and CSA. Indian Institutions and the Canadian Space Agency have contributed to the work presented in this paper. Several groups from ISAC (ISRO), Bengaluru, and IISU (ISRO), Trivandrum have contributed to the design, fabrication, and testing of the payload. The Mission Group (ISAC) and ISTRAC (ISAC) continue to provide support in making observations with, and reception and initial processing of the data. We gratefully thank all the individuals involved in the various teams for providing their support to the project from the early stages of the design to launch and observations with it in the orbit. This work also uses observations taken by the 3D-HST Treasury Program (HST-GO-12177 and HST-GO-12328) with the NASA/ESA Hubble Space Telescope, which is operated by the Association of Universities for Research in Astronomy, Inc., under
NASA contract NAS5-26555. This research made use of Matplotlib \citep{matplotlib2007}, Astropy \citep{astropy2013,astropy2018}, community-developed core Python packages for Astronomy and SAOImageDS9 \citep{joye2003}. Finally, we thank the referee for valuable suggestions.

\software{SExtractor \citep{bertin1996}, SAOImageDS9 \citep{joye2003}, Matplotlib \citep{matplotlib2007}, Astropy \citep{astropy2013,astropy2018}}

%\bibliographystyle{aasjournal}
%\bibliography{reference}

\end{document}